\documentclass[twoside,12pt]{article}%
\usepackage{amssymb}
\usepackage{amsfonts}
\usepackage{amsmath}
\usepackage{graphicx}%
\providecommand{\U}[1]{\protect\rule{.1in}{.1in}}
\newtheorem{theorem}{Theorem}

\newtheorem{corollary}[theorem]{Corollary}

\newtheorem{definition}[theorem]{Definition}

\newtheorem{lemma}[theorem]{Lemma}

\newtheorem{remark}[theorem]{Remark}

\newenvironment{proof}[1][Proof]{\noindent\textbf{#1.} }{\ \rule{0.5em}{0.5em}}
\begin{document}

\title{\textbf{Analytical} \textbf{Blowup Solutions to the }$2$\textbf{-dimensional
Isothermal Euler-Poisson Equations of Gaseous Stars}}
\author{Y\textsc{uen} M\textsc{anwai\thanks{E-mail address: nevetsyuen@hotmail.com }}\\\textit{Department of Mathematics and Statistics, Hang Seng School of
Commerce,}\\\textit{Hang Shin Link, Siu Lek Yuen, Shatin,New Territories, Hong Kong}}
\date{Revised 28-Sept-2007}
\maketitle

\begin{abstract}
We study the Euler-Poisson equations of describing the evolution of the
gaseous star in astrophysics. Firstly, we construct a family of analytical
blowup solutions for the isothermal case in $R^{2}$. Furthermore the blowup
rate of the above solutions is also studied and some remarks about the
applicability of such solutions to the Navier-Stokes-Poisson equations and the
drift-diffusion model in semiconductors are included. Finally, for the
isothermal case $(\gamma=1)$, the result of Makino and Perthame for the tame
solutions is extended to show that the life span of such solutions must be
finite if the initial data is with compact support.

\end{abstract}

\section{Introduction}

The evolution of a self-gravitating fluid such as gaseous stars can be
formulated by the Euler-Poisson equation of the following form:
\begin{equation}%
\begin{array}
[c]{rl}%
{\normalsize \rho}_{t}{\normalsize +\nabla\cdot(\rho u)} & {\normalsize =}%
{\normalsize 0,}\\
{\normalsize (\rho u)}_{t}{\normalsize +\nabla\cdot(\rho u\otimes u)+\nabla P}
& {\normalsize =}{\normalsize -\rho\nabla\Phi,}\\
S_{t}+u\cdot\nabla S & =0,\\
{\normalsize \Delta\Phi(t,x)} & {\normalsize =\alpha(N)g}{\normalsize \rho,}%
\end{array}
\label{Euler-Poisson}%
\end{equation}
where $\alpha(N)$ is a constant related to the unit ball in $R^{N}$:
$\alpha(1)=2$; $\alpha(2)=2\pi$; For simplicity, we take the constant term
$g=1$. For $N\geq3,$%
\[
\alpha(N)=N(N-2)V(N)=N(N-2)\frac{\pi^{N/2}}{\Gamma(N/2+1)},
\]
where $V(N)$ is the volume of the unit ball in $R^{N}$ and $\Gamma$ is a Gamma
function. As usual, $\rho=\rho(t,x)$, $u=u(t,x)\in\mathbf{R}^{N}$ and $S(t,x)$
are the density, the velocity and the entropy respectively. $P=P(\rho)$\ is
the pressure. In the above system, the self-gravitational potential field
$\Phi=\Phi(t,x)$\ is determined by the density $\rho$ through the Poisson
equation. The equations (\ref{Euler-Poisson})$_{1}$ and (\ref{Euler-Poisson}%
)$_{2}$ are the compressible Euler equation with forcing term. The equation
(\ref{Euler-Poisson})$_{3}$ is the Poisson equation through which the
gravitational potential is determined by the density distribution of the gas
itself. Thus, we called the system (\ref{Euler-Poisson}) the Euler-Poisson
equations. Here, the viscosity term does not appear, that is, the viscous
effect is neglected.\ In this case, the equations can be viewed as a prefect
gas model. For $N=3$, (\ref{Euler-Poisson}) is a classical (nonrelativistic)
description of a galaxy, in astrophysics. See \cite{C}, \cite{M1} for a detail
about the system.

If we take $S(t,x)=\ln K$, for some fixed $K>0$, we have a $\gamma$-law on the
pressure $P(\rho)$, i.e.%
\begin{equation}
{\normalsize P}\left(  \rho\right)  {\normalsize =K\rho}^{\gamma}\doteq
\frac{{\normalsize \rho}^{\gamma}}{\gamma}, \label{gamma}%
\end{equation}
which is a commonly the hypothesis. The constant $\gamma=c_{P}/c_{v}\geq1$,
where $c_{P}$, $c_{v}$\ are the specific heats per unit mass under constant
pressure and constant volume respectively, is the ratio of the specific heats,
that is, the adiabatic exponent in (\ref{gamma}). In particular, the fluid is
called isothermal if $\gamma=1$. It can be used for constructing models with
non-degenerate isothermal cores, which have a role in connection with the
so-called Schonberg-Chandrasekhar limit \cite{KW}.

For the physical dimension $N=3$, we are interested in the hydrostatic
equilibrium specified by $u=0,S=\ln K$. According to \cite{C}, the ratio
between the core density $\rho(0)$ and the mean density $\overset{\_}{\rho}$
for $6/5<\gamma<2$\ is given by%
\[
\frac{\overset{\_}{\rho}}{\rho(0)}=\left(  \frac{-3}{z}\overset{\cdot}%
{y}\left(  z\right)  \right)  _{z=z_{0}}%
\]
where $y$\ is the solution of the Lane-Emden equation with $n=1/(\gamma-1)$,%
\[
\overset{\cdot\cdot}{y}+\frac{2}{z}\overset{\cdot}{y}+y^{n}=0,\text{
}y(0)=\alpha>0,\text{ }\overset{\cdot}{y}(0)=0,\text{ }n=\frac{1}{\gamma-1},
\]
and $z_{0}$\ is the first zero of $y(z_{0})=0$. We can solve the Lane-Emden
equation analytically for%
\[
y_{anal}(z)\doteq\left\{
\begin{array}
[c]{ll}%
1-\frac{1}{6}z^{2}, & n=0;\\
\frac{\sin z}{z}, & n=1;\\
\frac{1}{\sqrt{1+z^{2}/3}}, & n=5,
\end{array}
\right.
\]
and for the other values, only numerical values can be obtained. It can be
shown that for $n<5$, the radius of polytropic models is finite; for $n\geq5$,
the radius is infinite.

Gambin \cite{G} and Bezard \cite{B} obtained the existence results about the
explicitly stationary solution $\left(  u=0\right)  $ for $\gamma=6/5:$%
\begin{equation}
\rho=\left(  \frac{3KA^{2}}{2\pi}\right)  ^{5/4}\left(  1+A^{2}r^{2}\right)
^{-5/2}. \label{stationsoluionr=6/5}%
\end{equation}
The Poisson equation (\ref{Euler-Poisson})$_{3}$ can be solved as%
\[
{\normalsize \Phi(t,x)=}\int_{R^{N}}G(x-y)\rho(t,y){\normalsize dy,}%
\]
where $G$ is the Green's function for the Poisson equation in the
$N$-dimensional spaces defined by
\[
G(x)\doteq\left\{
\begin{array}
[c]{ll}%
|x|, & N=1;\\
\log|x|, & N=2;\\
\frac{-1}{|x|^{N-2}}, & N\geq3.
\end{array}
\right.
\]
In the following, we always seek solutions in spherical symmetry. Thus, the
Poisson equation (\ref{Euler-Poisson})$_{3}$ is transformed to%
\[
{\normalsize r^{N-1}\Phi}_{rr}\left(  {\normalsize t,x}\right)  +\left(
N-1\right)  r^{N-2}\Phi_{r}{\normalsize =}\alpha\left(  N\right)
{\normalsize \rho r^{N-1},}%
\]%
\[
\Phi_{r}=\frac{\alpha\left(  N\right)  }{r^{N-1}}\int_{0}^{r}\rho
(t,s)s^{N-1}ds.
\]

\begin{definition}
[Blowup]We say a solution blows up if one of the following conditions is
satisfied:\newline(1)The solution becomes infinitely large at some point $x$
and some finite time $T$;\newline(2)The derivative of the solution becomes
infinitely large at some point $x$ and some finite time $T$.
\end{definition}

In this paper, we concern about blowup solutions for the $N$-dimensional
isothermal Euler-Poisson equations, which may describe the phenomenon called
the core collapsing in physics. And our aim is to construct a family of such
blowup solutions to it.

Historically in astrophysics, Goldreich and Weber \cite{GW} constructed the
analytical blowup solution (collapsing) solution of the $3$-dimensional
Euler-Poisson equation for $\gamma=4/3$ for the non-rotating gas spheres.
After that, Makino \cite{M1} obtained the rigorously mathematical proof of the
existence of such kind of blowup solutions. And in \cite{DXY}, we find the
extension of the above blowup solutions to the case $N\geq3$ and
$\gamma=(2N-2)/N$. In \cite{Y}, the solutions with a from is written as%
\begin{equation}
\left\{
\begin{array}
[c]{c}%
\rho(t,r)=\left\{
\begin{array}
[c]{c}%
\frac{1}{a(t)^{N}}y(\frac{r}{a(t)})^{N/(N-2)},\text{ for }r<a(t)Z_{\mu};\\
0,\text{ for }a(t)Z_{\mu}\leq r.
\end{array}
\right.  \text{, }{\normalsize u(t,r)=}\frac{\overset{\cdot}{a}(t)}%
{a(t)}{\normalsize r,}\text{ }S(t,r)=LnK\\
\overset{\cdot\cdot}{a}(t){\normalsize =}-\frac{\lambda}{a(t)^{N-1}},\text{
}{\normalsize a(0)=a}_{0}>0{\normalsize ,}\text{ }\overset{\cdot}%
{a}(0){\normalsize =a}_{1},\\
\overset{\cdot\cdot}{y}(z){\normalsize +}\frac{N-1}{z}\overset{\cdot}%
{y}(z){\normalsize +}\frac{N(N-2)^{2}V(N)}{(2N-2)K}{\normalsize y(z)}%
^{N/(N-2)}{\normalsize =\mu,}\text{ }y(0)=\alpha>0,\text{ }\overset{\cdot}%
{y}(0)=0,
\end{array}
\right.  \label{solution2}%
\end{equation}
where $\mu=[N(N-2)\lambda]/(2N-2)K$ and the finite $Z_{\mu}$ is the first zero
of $y(z)$. And no other analytical blowup solution for the Euler-Poisson
equation has been obtained. Through there are a lot of numerical simulation
results on the Euler-Poisson equation.

In section 2, we obtain the blowup solutions for the Euler-Poisson equation in
spherical symmetry in the $2$-dimensional case,%
\begin{equation}%
\begin{array}
[c]{rl}%
\rho_{t}+u\rho_{r}+\rho u_{r}+{\normalsize \frac{1}{r}\rho u} &
{\normalsize =0,}\\
\rho\left(  u_{t}+uu_{r}\right)  +K\rho_{r} & {\normalsize =-}\frac{2\pi\rho
}{r}\int_{0}^{r}\rho(t,s)sds,
\end{array}
\label{gamma=1}%
\end{equation}
in the form of the following theorem.

\begin{theorem}
\label{thm:1}For the $2$-dimensional Euler-Poisson equations in radial
symmetry, (\ref{gamma=1}), there exists a family of solutions,%
\begin{equation}
\left\{
\begin{array}
[c]{c}%
\rho(t,r)=\frac{1}{a(t)^{2}}e^{y(r/a(t))}\text{, }{\normalsize u(t,r)=}%
\frac{\overset{\cdot}{a}(t)}{a(t)}{\normalsize r;}\\
\overset{\cdot\cdot}{a}(t){\normalsize =}-\frac{\lambda}{a(t)},\text{
}{\normalsize a(0)=a}_{0}>0{\normalsize ,}\text{ }\overset{\cdot}%
{a}(0){\normalsize =a}_{1};\\
\overset{\cdot\cdot}{y}(x){\normalsize +}\frac{1}{x}\overset{\cdot}%
{y}(x){\normalsize +\frac{\alpha(N)}{K}e}^{y(x)}{\normalsize =\mu,}\text{
}y(0)=\alpha,\text{ }\overset{\cdot}{y}(0)=0,
\end{array}
\right.  \label{solution1}%
\end{equation}
where $K>0$, $\mu=2\lambda/K$ with a sufficiently small $\lambda$ and $\alpha$
are constants.\newline(1)When $\lambda>0$, the solutions blow up in a finite
time $T$;\newline(2)When $\lambda=0$, if $a_{1}<0$, the solutions blow up at
$t=-a_{0}/a_{1}$.
\end{theorem}

The blowup rate of the solution (\ref{solution1}) is discussed in section 3.
Actually, we have

\begin{theorem}
\label{thm:2}The blowup rate of the solution (\ref{solution1})\textbf{ }is,%
\[
\underset{t\rightarrow T}{\lim}\rho(t,0)\left(  T-t\right)  ^{\eta}\geq O(1),
\]
with $\eta<2$.
\end{theorem}

In the last section, we can rewrite (\ref{Euler-Poisson}) in scalar form,%
\begin{equation}%
\begin{array}
[c]{rl}%
\frac{\partial\rho}{\partial t}+\sum_{k=1}^{N}u_{k}\frac{\partial\rho
}{\partial x_{k}}+\rho\sum_{k=1}^{N}\frac{\partial u_{k}}{\partial x_{k}} &
{\normalsize =}{\normalsize 0,}\\
\rho\left(  \frac{\partial u_{i}}{\partial t}+\sum_{k=1}^{N}u_{k}%
\frac{\partial u_{i}}{\partial x_{k}}\right)  +\frac{\partial P}{\partial
x_{i}}+\rho\frac{\partial\Phi}{\partial x_{i}} & {\normalsize =0}\text{, for
}i=1,2,...N.
\end{array}
\label{Tame}%
\end{equation}
The tame solution is introduced in here.

\begin{definition}
A solution $(\rho,u)$ for (\ref{Tame}) is called "tame" solution
if\newline(1)$(\rho,u)\in C^{1}([0,T)\times R^{N}),$ $\rho\geq0,$ $\rho(t)$
has compact support, and\newline(2)$\rho^{(\gamma-1)/2}\in C^{1}([0,T)\times
R^{N}),$ $u\in C([0,T);B^{0})$, and
\begin{equation}
\frac{\partial u_{i}}{\partial t}+\sum\nolimits_{k=1}^{N}u_{k}\frac{\partial
u_{i}}{\partial x_{k}}+\frac{\partial\Phi}{\partial x_{i}}=0,\text{ for
}i=1,2,...N \label{deftame2}%
\end{equation}
holds on the exterior of the support of $\rho$.
\end{definition}

If only the condition (1) is assumed, the solution is called "classical". For
$\gamma>1$, under the transformation%
\[
w=\frac{2\sqrt{K\gamma}}{\gamma-1}\rho^{(\gamma-1)/2},
\]
Makino and Perthame \cite{MP} showed that the life span of tame solution of
the Euler-Poisson equation in spherical symmetry is finite. As we are
interested in non-global existence for the isothermal case $(\gamma=1)$ in $2$
dimension, i.e.%
\begin{equation}%
\begin{array}
[c]{rl}%
\rho_{t}+u\rho_{r}+\rho u_{r}+{\normalsize \frac{1}{r}\rho u} &
{\normalsize =0,}\\
\rho\left(  u_{t}+uu_{r}\right)  +K\rho_{r} & {\normalsize =-}\frac{2\pi\rho
}{r}\int_{0}^{r}\rho(t,s)sds,
\end{array}
\label{gamma1}%
\end{equation}
the result of Makino and Perthame \cite{MP} for the tame solutions is extended
as the following theorem.

\begin{theorem}
\label{Thmtamesolution}Let $(\rho(t),u(t))$ be a radially symmetric tame
solution of (\ref{gamma1}) on $0\leq t<T$. If the support of $(\rho(0),u(0))$
is compact and $\rho(0)$ is not identically equal to zero, then $T$ must be finite.
\end{theorem}

\section{Separable Blowup Solutions}

In this section, before presenting the proof of Theorem \ref{thm:1}, we
prepare some lemmas. First, we obtain a general class of solutions for the
continuity equation of mass in radial symmetry (\ref{gamma=1})$_{1}$.

\begin{lemma}
\label{lem:generalsolutionformasseq}For the equation of conservation of mass
in radial symmetry
\begin{equation}
\rho_{t}+u\rho_{r}+\rho u_{r}+\frac{1}{r}\rho u=0,
\label{massequationspherical}%
\end{equation}
there exist solutions,%
\begin{equation}
\rho(t,r)=\frac{f(r/a(t))}{a(t)^{2}},\text{ }{\normalsize u(t,r)=}%
\frac{\overset{\cdot}{a}(t)}{a(t)}{\normalsize r,}
\label{generalsolutionformassequation}%
\end{equation}
with the form with $f\geq0\in C^{1}$ and $a(t)>0\in C^{1}.$
\end{lemma}

\begin{proof}
We just plug (\ref{generalsolutionformassequation}) into
(\ref{massequationspherical}). Then
\begin{align*}
&  \rho_{t}+u\rho_{r}+\rho u_{r}+\frac{1}{r}\rho u\\
&  =\frac{-2\overset{\cdot}{a}(t)f(r/a(t))}{a(t)^{3}}-\frac{\overset{\cdot}%
{a}(t)r\overset{\cdot}{f}(r/a(t))}{a(t)^{4}}\\
&  +\frac{\overset{\cdot}{a}(t)r}{a(t)}\frac{\overset{\cdot}{f}(r/a(t))}%
{a(t)^{3}}+\frac{f(r/a(t))}{a(t)^{2}}\frac{\overset{\cdot}{a}(t)}{a(t)}%
+\frac{1}{r}\frac{f(r/a(t))}{a(t)^{2}}\frac{\overset{\cdot}{a}(t)}{a(t)}r\\
&  =0.
\end{align*}
The proof is completed.
\end{proof}

Secondly, we obtain an estimate to the equation,%
\begin{equation}%
\begin{array}
[c]{l}%
\ddot{a}(t)=-\frac{\lambda}{a(t)},\\[3mm]%
a(0)=a_{0}>0,\ \dot{a}(0)=a_{1}.
\end{array}
\label{Lane-Emden}%
\end{equation}

\begin{lemma}
$\label{lemma1}$For the Emden equation (\ref{Lane-Emden}), we have,\newline%
(1)If $\lambda>0$, there exists a finite time $T_{-}<+\infty$ such that
$a(T_{-})=0$;\newline(2)If $\lambda=0$, it holds that for any $t\geq0$,%
\[
a(t)=a_{0}+a_{1}t.
\]

\end{lemma}

\begin{proof}
(1) By integrating (\ref{Lane-Emden}), we have%
\begin{equation}
0\leq\frac{1}{2}\overset{\cdot}{a}(t)^{2}=-\lambda\ln a(t)+\theta\label{eq1}%
\end{equation}
where $\theta=\lambda\ln a_{0}+\frac{1}{2}a_{1}^{2}.$\newline From
(\ref{eq1}), we get,%
\[
a(t)\leq e^{\theta/\lambda}.
\]
If the statement (1) is not true, we have%
\[
0<a(t)\leq e^{\theta/\lambda},\text{ for all }t\geq0.
\]
But since
\[
\ddot{a}(t)=-\frac{\lambda}{a(t)}\leq\frac{-\lambda}{e^{\theta/\lambda}},
\]
we integrate this twice to deduce%
\[
a(t)\leq\int_{0}^{t}\int_{0}^{\tau}\frac{-\lambda}{e^{\theta/\lambda}}%
dsd\tau+C_{1}t+C_{0}=\frac{-\lambda t^{2}}{2e^{\theta/\lambda}}+C_{1}t+C_{0}.
\]
By taking $t$ large enough, we get%
\[
a(t)<0.
\]
As a contradiction is met, the statement (1) is true. It is trivial to verify (2).
\end{proof}

\begin{remark}
The modified systems of the above Emden equation may be referred more
detailedly in \cite{C}, \cite{DXY},\cite{Y}.
\end{remark}

\begin{lemma}
\label{lemma2}There exists a sufficiently small $x_{0}>0$, such that the
equation%
\begin{equation}
\left\{
\begin{array}
[c]{c}%
\overset{\cdot\cdot}{y}(x){\normalsize +}\frac{1}{x}\overset{\cdot}%
{y}(x){\normalsize +\sigma e}^{y(x)}{\normalsize =\mu,}\\
y(0)=\alpha,\text{ }\overset{\cdot}{y}(0)=0,
\end{array}
\right.  \label{SecondorderElliptic}%
\end{equation}
where $\sigma>0$, $\mu$, and $\alpha$ are constants, has a solution
$y=y(x,\mu)\in C^{2}[0,x_{0}]$.
\end{lemma}

\begin{proof}
The lemma can be proved by the fixed point theorem. Multiply
(\ref{SecondorderElliptic}) by $x$, to\ give%
\[
\frac{d}{dx}\left(  x\overset{\cdot}{y}(x)\right)  =x\left(  \mu-\sigma
e^{y(x)}\right)  .
\]
Notice $\overset{\cdot}{y}(0)=0$,\ we have%
\[
\overset{\cdot}{y}(x)=\frac{1}{x}\int_{0}^{x}s(\mu-\sigma e^{y(s)})ds.
\]
By using $y(0)=\alpha$, (\ref{SecondorderElliptic}) is reduced to%
\[
\overset{\cdot}{y}(x)=\frac{1}{x}\int_{0}^{x}s(\mu-\sigma e^{y(s)})ds,\text{
}y(0)=\alpha.
\]
Set%
\[
{\normalsize f(x,y(x))=}\frac{1}{x}\int_{0}^{x}{\normalsize s(\mu-}\sigma
e^{y(s)}{\normalsize )ds.}%
\]
then for any $x_{0}>0$, we get $f\in C^{1}[0,$ $x_{0}]$. and for any $y_{1,}$
$y_{2}\in C^{2}[0,$ $x_{0}]$, we have,%
\[
\left\vert f(x,y_{1}(x))-f(x,y_{2}(x))\right\vert =\frac{\sigma\left\vert
\int_{0}^{x}s(e^{y_{2}(s)}-e^{y_{1}(s)})ds\right\vert }{x}.
\]
As $e^{y}$ is a $C^{1}$ function of $y$, we can show that the function $e^{y}%
$, is Lipschitz-continuous. And we get,%
\begin{align*}
&  \left\vert f(x,y_{1}(x))-f(x,y_{2}(x)\right\vert \\
&  =\frac{O(1)\int_{0}^{x}s\left\vert \left(  y_{2}(s\right)  -y_{1}%
(s)\right\vert ds}{x}\\
&  \leq O(1)x_{0}\underset{0\leq x\leq x_{0}}{\sup}\left\vert y_{1}%
(s)-y_{2}(s)\right\vert ,
\end{align*}
where $\tau\in\lbrack0,$ $x]\subseteq\lbrack0,$ $x_{0}]$. Let%
\[
{\normalsize Ty(x)=\alpha+}\int_{0}^{x}{\normalsize f(s,y(s))ds.}%
\]
We have $Ty\in C[0,$ $x_{0}]$\ and%
\begin{align*}
&  \left\vert Ty_{1}(x)-Ty_{2}(x)\right\vert \\
&  =\left\vert \int_{0}^{x}f(s,y_{1}(s))ds-\int_{0}^{x}f(s,y_{2}%
(s))ds\right\vert \\
&  \leq O(1)x_{0}\underset{0\leq x\leq x_{0}}{\sup}\left\vert y(x)_{1}%
-y(x)_{2}\right\vert .
\end{align*}
By choosing $x_{0}>0$ to be a sufficiently small number, such that
$O(1)x_{0}<1$, this shows that the mapping $T:C[0,$ $X_{0}]\rightarrow C[0,$
$x_{0}]$, is a contraction with the sup-norm. By the fixed point theorem,
there exists a unique $y(x)\in C[0,$ $x_{0}],$\ such that $Ty(x)=y(x)$. The
proof is completed.
\end{proof}

\begin{lemma}
\label{lemma3}The equation,%
\begin{equation}
\left\{
\begin{array}
[c]{c}%
\overset{\cdot\cdot}{y}(x){\normalsize +}\frac{1}{x}\overset{\cdot}%
{y}(x){\normalsize +\sigma e}^{y(x)}{\normalsize =0,}\\
y(0)=\alpha,\text{ }\overset{\cdot}{y}(0)=0,
\end{array}
\right.  \label{Elliptic1}%
\end{equation}
where $\sigma>0$ and $\alpha$ are constants, has a solution in $[0,$
$+\infty)$ and $\underset{x\rightarrow+\infty}{\lim}y(x)=-\infty$.
\end{lemma}

\begin{proof}
By integrating (\ref{Elliptic1}), we have,%
\begin{equation}
\overset{\cdot}{y}(x)=-\frac{\sigma}{x}\int_{0}^{x}se^{y(s)}ds\leq
0.\label{lemma3eq1}%
\end{equation}
Thus, for $0<x<x_{0}$, $y(x)$ has a uniform lower upper bound
\[
y(x)\leq y(0)=\alpha.
\]
As we obtained he local existence in Lemma \ref{lemma2}, there are two
possibilities:\newline(1)$y(x)$ only exists in some finite interval
$[0,x_{0}]$: (1a)$\underset{x\rightarrow x_{0-}}{\lim}y(x)=-\infty$;
(1b)$y(x)$ has an uniformly lower bound, i.e. $y(x)\geq\alpha_{0}$ for some
constant $\alpha_{0}.$\newline(2)$y(x)$ exists in $[0,$ $+\infty)$:
(2a)$\underset{x\rightarrow+\infty}{\lim}y(x)=-\infty$; (2b)$y(x)$ has an
uniformly lower bound, i.e. $y(x)\geq\alpha$ for some constant $\alpha_{0}%
$.\newline We claim that possibility (1) doesn't exist. We need to reject (1b)
first: If the statement (1b) is true, (\ref{lemma3eq1}) becomes%
\begin{equation}
-\sigma xe^{\alpha}=-\frac{\sigma}{x}\int_{0}^{x}xe^{\alpha}ds\leq
\overset{\cdot}{y}(x).\label{possible1}%
\end{equation}
Thus, $\overset{\cdot}{y}(x)$ is bounded in $[0,x_{0}]$. Therefore, we can use
the fixed point theorem again to obtain a large domain of existence, such that
$[0,x_{0}+\delta]$ for some positive number $\delta$. There is a
contradiction. Therefore, (1b) is rejected.\newline Next, we do not accept
(1a) because of the following reason: It is impossible that $\underset
{x\rightarrow x_{0-}}{\lim}y(x)=-\infty$, as from (\ref{possible1}),
$\overset{\cdot}{y}(x)$ has a lower bound in $[0,$ $x_{0}]$:%
\begin{equation}
-\sigma x_{0}e^{\alpha}\leq\overset{\cdot}{y}(x).\label{lemma3eq2}%
\end{equation}
Thus, (\ref{lemma3eq2}) becomes,
\begin{align*}
y(x_{0}) &  =y(0)+\int_{0}^{x_{0}}\overset{\cdot}{y}(x)dx\\
&  \geq\alpha-\int_{0}^{x_{0}}\sigma x_{0}e^{\alpha}dx\\
&  =\alpha-\sigma x_{0}^{2}e^{\alpha}.
\end{align*}
Since $y(x)$ is bounded below in $[0,$ $x_{0}]$, it contracts the statement
(1a), such that $\underset{x\rightarrow x_{0-}}{\lim}y(x)=-\infty$. So, we can
exclude the possibility (1).\newline We claim that the possibility (2b)
doesn't exist. It is because
\[
\overset{\cdot}{y}(x)=-\frac{\sigma}{x}\int_{0}^{x}se^{y(s)}ds\leq
-\frac{\sigma}{x}\int_{0}^{x}e^{\alpha_{0}}sds=-\frac{\sigma e^{a_{0}}x}{2}.
\]
Then, we have,%
\begin{equation}
y(x)\leq\alpha-\frac{\sigma e^{a_{0}}}{4}x^{2}.\label{lemma3eq3}%
\end{equation}
By letting $x\rightarrow\infty$, (\ref{lemma3eq3}) turns out to be,
\[
y(x)=-\infty.
\]
Since a contradiction is established, we exclude the possibility (2b). Thus,
the equation (\ref{Elliptic1}) exists in $[0,$ $+\infty)$ and $\underset
{x\rightarrow+\infty}{\lim}y(x)=-\infty$. This completes the proof.
\end{proof}

Now transferring (\ref{SecondorderElliptic}) to the first-order system,%
\[
\left\{
\begin{array}
[c]{c}%
\frac{dy}{dx}=\overset{\cdot}{y},\\
\frac{d\overset{\cdot}{y}}{dx}=-\frac{1}{x}\overset{\cdot}{y}-\sigma e^{y}%
+\mu,
\end{array}
\right.
\]
we consider the system along the solution curve $y=y(x,0)$, $\overset{\cdot
}{y}=\overset{\cdot}{y}(x,0)$, $0<x_{0}\leq x<+\infty$. Since the right-hand
side of this system is continuously differentiable in $x,$ $y,$ $\overset
{\cdot}{y}$ and $\mu$. We apply the comparison theorem (Theorem 7.4) in
\cite{CL}, to obtain the following lemma.

\begin{lemma}
\label{lemma4}For a sufficiently small $\epsilon$ such that $\left\vert
\mu\right\vert <\epsilon$, the equation,%
\[
\left\{
\begin{array}
[c]{c}%
\overset{\cdot\cdot}{y}(x){\normalsize +}\frac{1}{x}\overset{\cdot}%
{y}(x){\normalsize +\sigma e}^{y(x)}{\normalsize =\mu,}\\
y(0)=\alpha,\overset{\cdot}{y}(0)=0,
\end{array}
\right.
\]
exists in $[0,$ $\infty),$ and $\underset{x\rightarrow+\infty}{\lim
}y(x)=-\infty$.
\end{lemma}

Now, we give the proof of Theorem \ref{thm:1}.

\begin{proof}
[Proof of Theorem 2]From Lemma \ref{lemma4}, we easily get that
(\ref{solution1}) satisfy (\ref{gamma=1})$_{1}$. For the momentum equation
(\ref{gamma=1})$_{2}$, we get,%
\begin{align*}
&  \rho(u_{t}+uu_{r})+K\rho_{r}+\frac{2\pi\rho}{r}%
{\displaystyle\int\limits_{0}^{r}}
\rho(t,s)sds\\
&  =\frac{\rho}{a(t)}\left[  \overset{\cdot\cdot}{a}(t)r{\normalsize +}%
K\overset{\cdot}{y}(\frac{r}{a(t)})+\frac{2\pi}{ra(t)}%
{\displaystyle\int\limits_{0}^{r}}
e^{y\left(  s/a(t)\right)  }sds\right] \\
&  =\frac{\rho}{a(t)}\left[  -\frac{\lambda r}{a(t)}+K\overset{\cdot}{y}%
(\frac{r}{a(t)})+\frac{2\pi}{(\frac{r}{a(t)})}%
{\displaystyle\int\limits_{0}^{r/a(t)}}
e^{y\left(  s\right)  }sds\right] \\
&  =\frac{\rho}{a(t)}Q\left(  \frac{r}{a(t)}\right)  .
\end{align*}
Here, we use the property of $a(t)$:%
\[
\overset{\cdot\cdot}{a}(t){\normalsize =}-\frac{\lambda}{a(t)},
\]
and denote%
\[
Q(\frac{r}{a(t)})={\normalsize Q(x)=-\lambda x+}K\overset{\cdot}%
{y}(x){\normalsize +}\frac{2\pi}{x}%
{\displaystyle\int\limits_{0}^{x}}
e^{y\left(  s\right)  }sds{\normalsize .}%
\]
Differentiate $Q(x)$\ with respect to $x$,%
\begin{align*}
\overset{\cdot}{Q}(x)  &  =-{\normalsize \lambda+}K\overset{\cdot\cdot}%
{y}(x){\normalsize +}2\pi e^{y(x)}-\frac{2\pi}{x}%
{\displaystyle\int\limits_{0}^{x}}
e^{y(s)}{\normalsize sds}\\
&  =-{\normalsize \lambda+}K\left(  -\frac{1}{x}\overset{\cdot}{y}%
(x)+\mu\right)  -\frac{2\pi}{x}%
{\displaystyle\int\limits_{0}^{x}}
e^{y(s)}{\normalsize sds}\\
&  =-\frac{1}{x}\left(  \lambda x+K\overset{\cdot}{y}(x)-K\mu x+\frac{2\pi}{x}%
{\displaystyle\int\limits_{0}^{x}}
e^{y(s)}{\normalsize sds}\right) \\
&  =-\frac{1}{x}Q(x).
\end{align*}
The above result is due to the fact that we choose the following ordinary
differential equation:%
\[
\left\{
\begin{array}
[c]{c}%
\overset{\cdot\cdot}{y}(x){\normalsize +}\frac{1}{x}\overset{\cdot}%
{y}(x){\normalsize +\frac{2\pi}{K}e}^{y(x)}{\normalsize =\mu,}\text{ }%
\mu=\frac{2\lambda}{K},\\
{\normalsize y(0)=\alpha,}\text{ }\overset{\cdot}{y}(0){\normalsize =0.}%
\end{array}
\right.
\]
With $Q(0)=0$, this implies that $Q(x)=0$. By using Lemma \ref{lemma1} about
$a(t)$ and Lemma \ref{lemma3} about $y(x)$, we have shown that the family of
the solutions blows up in finite time $T$ under the prescribed conditions of
Theorem \ref{thm:1}. This completes the proof.
\end{proof}

\begin{remark}
Besides, the above solution requiring $y(0)>0$ is not necessary for
(\ref{solution1}) in contrast to (\ref{solution2}) in \cite{DXY}, \cite{M1}.
And the mass of the above solution is
\[
M=\int_{R^{2}}\rho(t,s)ds=\int_{0}^{+\infty}\frac{2\pi}{a(0)^{2}}%
e^{y(s/a(0))}sds=2\pi\int_{0}^{+\infty}e^{y(s)}sds,
\]
where $\eta(N)$ is the measure of a unit ball. Thus the mass of the solution
(\ref{solution1}) depends on the initial data $y(0)$. And it is not easy to
determine if the mass is finite or infinite with different datum. It is
different for the finite mass of the family of solutions (\ref{solution2}) for
$\gamma=4/3$\ and $N=3$ which is independent of $y(0)$ in \cite{DLYY}.
\end{remark}

\begin{remark}
Now, we consider the stationary solutions of (\ref{Euler-Poisson}), in
spherical symmetry, i.e.%
\begin{equation}
P_{r}=-\rho\Phi_{r}. \label{remarkeq1}%
\end{equation}
For $\gamma=1$, (\ref{remarkeq1}) becomes,%
\[
\rho=e^{-\Phi/K+C},
\]
where $C$ is a constant.\newline For our convenience, $C=0$ is chosen,%
\[
\rho=e^{-\Phi/K}.
\]
From the Poisson equation (\ref{Euler-Poisson})$_{3}$, we obtain,%
\begin{equation}
\Delta\Phi=\alpha(N)e^{-\Phi/K}. \label{remarkeq2}%
\end{equation}
As the solution is in radially symmetric, (\ref{remarkeq2}) becomes,%
\[
\Phi_{rr}+\frac{1}{r}\Phi_{r}-2\pi e^{-\Phi/K}=0.
\]
Under the transformation,
\[
y(x)=-\frac{\Phi(r)}{K},
\]
we have%
\begin{equation}
\overset{\cdot\cdot}{y}(x){\normalsize +}\frac{1}{x}\overset{\cdot}%
{y}(x){\normalsize +\frac{2\pi}{K}e}^{y(x)}{\normalsize =0.} \label{y(x)}%
\end{equation}

\end{remark}

Furthermore by setting $\lambda=0$ and $\overset{\cdot}{a}(0)=0$ in Theorem
\ref{thm:1}, we easily have the following corollary for the non-trivial
stationary solutions.

\begin{corollary}
\label{Cor1}For the 2-dimensional isothermal Euler-Poisson equation, there
exists a family of stationary solution, i.e.%
\begin{equation}
\left\{
\begin{array}
[c]{c}%
\rho(t,r)=\frac{1}{a^{2}}e^{^{y(r/a)}};\\
\overset{\cdot\cdot}{y}(x){\normalsize +}\frac{1}{x}\overset{\cdot}%
{y}(x){\normalsize +\frac{2\pi}{K}e}^{y(x)}{\normalsize =\mu,}\text{
}y(0)=\alpha,\text{ }\overset{\cdot}{y}(0)=0,
\end{array}
\right.  \label{stationary}%
\end{equation}
where $\alpha$ is an arbitrary positive constant.
\end{corollary}

\begin{remark}
Our blowup solutions only work for the $2$-dimensional case. But we do not
know what will happen\qquad\ after the critical time that the solutions blow
up.\newline If we consider the system of conservation laws with viscosity,
i.e. with $\nu>0$,
\[%
\begin{array}
[c]{rl}%
{\normalsize \rho}_{t}{\normalsize +\nabla\cdot(\rho u)} & {\normalsize =}%
{\normalsize 0,}\\
{\normalsize (\rho u)}_{t}{\normalsize +\nabla\cdot(\rho u\otimes u)+\nabla P}
& {\normalsize =}{\normalsize -\rho\nabla\Phi+\nu\Delta u,}\\
{\normalsize \Delta\Phi(t,x)} & {\normalsize =2\pi}{\normalsize \rho.}%
\end{array}
\]
We get the corresponding Navier-Stokes-Poisson equation. The equation
describes the situation closer to the real model compared with the
Euler-Poisson equations in gaseous stars. And our family of solutions is also
suitable for it. This is due to the fact in spherical symmetry for the vector
Laplacian in $u(t,r)$:%
\[
\Delta u=u_{rr}+\frac{1}{r}u_{r}-\frac{1}{r^{2}}u.
\]

\end{remark}

\begin{remark}
In \cite{Y}, we have the lemma to control the modified Emden equation with
$\beta>0$,%
\[
\overset{\cdot\cdot}{a}(t)+\beta a(t){\normalsize =}-\frac{\lambda}%
{a(t)},\text{ }{\normalsize a(0)=a}_{0}>0{\normalsize ,}\text{ }\overset
{\cdot}{a}(0){\normalsize =a}_{1}.
\]
The similar blowup results may be obtained for the $2$-dimensional
Euler-Poisson equation with frictional damping or Navier-Stokes-Poisson
equation with frictional damping or not, i.e.%
\[%
\begin{array}
[c]{rl}%
{\normalsize \rho}_{t}{\normalsize +\nabla\cdot(\rho u)} & {\normalsize =}%
{\normalsize 0,}\\
{\normalsize (\rho u)}_{t}{\normalsize +\nabla\cdot(\rho u\otimes u)+\nabla P}
& {\normalsize =-}{\normalsize \rho\nabla\Phi-\beta\rho u+\nu\Delta u,}\\
{\normalsize \Delta\Phi(t,x)} & {\normalsize =2\pi}{\normalsize \rho,}%
\end{array}
\]
where $\beta>0$ and $\nu\geq0$. The corresponding family of blowup solution is%
\begin{equation}
\left\{
\begin{array}
[c]{c}%
\rho(t,r)=\frac{1}{a(t)^{2}}e^{y(r/a(t))},{\normalsize u(t,r)=}\frac
{\overset{\cdot}{a}(t)}{a(t)}{\normalsize r;}\\
\overset{\cdot\cdot}{a}(t)+\beta a(t){\normalsize =}-\frac{\lambda}%
{a(t)},{\normalsize a(0)=a}_{0}>0{\normalsize ,}\overset{\cdot}{a}%
(0){\normalsize =a}_{1};\\
\overset{\cdot\cdot}{y}(x){\normalsize +}\frac{1}{x}\overset{\cdot}%
{y}(x){\normalsize +\frac{2\pi}{K}e}^{y(x)}{\normalsize =\mu,}y(0)=\alpha
,\overset{\cdot}{y}(0)=0,
\end{array}
\right.  \label{solution3}%
\end{equation}
where $\mu=2\lambda/K$ and $\alpha$ are constants.
\end{remark}

\begin{remark}
Besides, if we consider the drift-diffusion model in semiconductors,%
\[%
\begin{array}
[c]{rl}%
{\normalsize \rho}_{t}{\normalsize +\nabla\cdot(\rho u)} & {\normalsize =}%
{\normalsize 0,}\\
{\normalsize (\rho u)}_{t}{\normalsize +\nabla\cdot(\rho u\otimes u)+\nabla P}
& {\normalsize =+}{\normalsize \rho\nabla\Phi-\beta\rho u+\nu\Delta u,}\\
{\normalsize \Delta\Phi(t,x)} & {\normalsize =2\pi}{\normalsize \rho,}%
\end{array}
\]
the special solutions with infinite mass may be obtained as follows%
\begin{equation}
\left\{
\begin{array}
[c]{c}%
\rho(t,r)=\frac{1}{a(t)^{2}}e^{y(r/a(t))},{\normalsize u(t,r)=}\frac
{\overset{\cdot}{a}(t)}{a(t)}{\normalsize r;}\\
\overset{\cdot\cdot}{a}(t)+\beta a(t){\normalsize =}-\frac{\lambda}%
{a(t)},{\normalsize a(0)=a}_{0}>0{\normalsize ,}\overset{\cdot}{a}%
(0){\normalsize =a}_{1};\\
\overset{\cdot\cdot}{y}(x){\normalsize +}\frac{1}{x}\overset{\cdot}%
{y}(x)-{\normalsize \frac{2\pi}{K}e}^{y(x)}{\normalsize =\mu,}y(0)=\alpha
,\overset{\cdot}{y}(0)=0.
\end{array}
\right.
\end{equation}

\end{remark}

\section{Blowup Rate}

In this section, we present the confirmation of Theorem \ref{thm:2}. The
blowup rate of the constructed solutions (\ref{solution1}), for the
Euler-Poisson equation is studied. It is interesting to investigate how fast
the blowup solution tends to infinity as the time tends to the critical value
$T$.

\begin{proof}
[Proof of Theorem \ref{thm:2}]We choose a finite time $t$ to make $a(t)$
sufficiently small enough, such that $-\lambda\ln a+\theta>0$. From the lemma
\ref{lemma1}, it is clear for that there exists a finite time $t_{0}>0$ such
that $\overset{\cdot}{a}(t)<0$ for $t>t_{0}$. Next, since%
\[
\ddot{a}(t)=-\frac{\lambda}{a(t)}<0,
\]
when $t>t_{0}+\epsilon=t_{1}$, this means there exists a constant $\epsilon
>0$, such that%
\[
a(t)\leq e^{\theta/\lambda}-\delta,
\]
where $\delta$ is a sufficiently small positive constant, we have a smaller
upper bound of $a(t)$. On the other hand, as%
\[
\overset{\cdot}{a}=-\sqrt{-2\lambda\ln a+\theta}.
\]
It is easy to see that,
\begin{equation}
t=t_{1}+\int_{t_{1}}^{t}ds=t_{1}-\int_{a(t_{1})}^{a(t)}\frac{da}%
{\sqrt{-2\lambda\ln a+\theta}}\geq t_{1}+\int_{0}^{a(t)}\frac{O(1)da}%
{\sqrt{-\ln a}}. \label{emdeneq1}%
\end{equation}
We denote $\delta$ as a sufficiently small positive number, such that
$x\in(0,\delta],$%
\[
-\ln x\leq\frac{1}{x^{S}},
\]
where the constant $S>0$. Then, (\ref{emdeneq1}) becomes,%
\[
T-t_{1}\geq\int_{0}^{a(t)}\frac{O(1)da}{\sqrt{1/a^{S}}}=O(1)a(t)^{S/2+1}.
\]
By letting $t_{1}\rightarrow T_{-}$, we get,
\[
\underset{t_{1}\rightarrow T_{-}}{\lim}a(t)\left(  T-t_{1}\right)
^{-2/(S+2)}\leq O(1).
\]
As $y(0)=\alpha$, we can estimate the blowup rate at the origin by%
\[
\underset{t\rightarrow T_{-}}{\lim}\rho(t,0)\left(  T-t\right)  ^{4/(S+2)}\geq
O(1).
\]
Thus, we obtain,
\[
\underset{t\rightarrow T_{-}}{\lim}\rho(t,0)\left(  T-t\right)  ^{\eta}\geq
O(1),
\]
with $\eta<2$, and complete the proof.
\end{proof}

\section{Non-global Existence for Tame Solutions}

In this section, we present the proof for Theorem \ref{Thmtamesolution}. The
technique to Makino and Perthame's \cite{MP} is similar.

\begin{proof}
[Proof of Theorem \ref{Thmtamesolution} ]We denote the radius of the compact
support of the tame solution by $R(t)$. Under the assumption that the initial
data has a compact support, i.e. $\rho(0,r)=u(0,r)=0$ for $r\geq R(0)$, from
(\ref{gamma1})$_{2}$, we must have
\[
K\rho_{r}=0.
\]
And the equation (\ref{gamma1})$_{2}$ becomes
\[
u_{t}+uu_{r}={\normalsize -}\frac{2\pi}{r}\int_{0}^{r}\rho(t,s)sds,
\]

along the curve $r=r(t;0,y)$ where $r=r(t;t_{0},r_{0})$ is the solution of the
characteristic equation outside the support of the tame solution, i.e.%
\[
\frac{dr}{dt}=u(t,r),\text{ }r\left\vert _{t=t_{0}}\right.  =r_{0}.
\]
Clearly we have for $y\geq R(t)$,%
\[
\frac{d}{dt}u(t,r(t;0,y))=u_{t}+uu_{r}=-\Phi_{r}=G(r)\doteq-\frac{2\pi}{r}%
\int_{0}^{r}\rho(t,s)sds\leq0.
\]
Therefore, we get%
\[
\frac{d}{dt}r(t;0,y)=u(t,r(t;0,y))=\int_{0}^{t}G(\tau,r(\tau;0,y))d\tau\leq0,
\]
as $u(0,y)=0$. Hence $r(t;0,y)\leq y$ for $0\leq t<T$, $R(t)\leq y.$ For
$\rho(t)\equiv0$ outside the compact support of the tame solution, this
implies%
\[
G(t,r(t,0,R(0)))=-\frac{M}{r(t,0,R(0))},
\]
where%
\[
M=\int_{R^{2}}\rho(t,x)dx=2\pi\int_{0}^{\infty}\rho(t,s)sds,
\]
is the total mass and independent of $t$. And If $\rho(0)$ is not identically
equal to zero, as the solution belongs $C^{1}$, we have $M>0$. Therefore
$r=r(t;0,R(0))$ satisfies the equation of free fall%
\begin{equation}
\frac{d^{2}r}{dt^{2}}=-\frac{M}{r},r(0)=R(0),\overset{\cdot}{r}(0)=0.
\label{secondode}%
\end{equation}
From this it follows
\[
0\leq r\leq R(0)-\frac{M}{2R(0)}t^{2}.
\]
This implies
\[
T\leq\sqrt{2R(0)^{2}/M}.
\]
The proof is completed.
\end{proof}

\end{document}